\documentclass[a4paper,3p,sort&compress,11pt,times]{elsarticle}

\usepackage{graphicx}
\usepackage{enumerate}
\usepackage{amsmath,amssymb}
\usepackage{url}
\usepackage{booktabs}
\usepackage{multirow}
\usepackage{xspace}
\usepackage{color}

\usepackage[pagebackref=false,breaklinks=true,colorlinks,bookmarks=false,linkcolor=blue,citecolor=blue]{hyperref}

\journal{Pattern Recognition}

\def\Vec#1{{\boldsymbol{#1}}}
\def\Mat#1{{\boldsymbol{#1}}}

\def\argmin#1#2{{\underset{#1}{\operatorname{arg\,min}}~#2}}
\def\etal{et~al.}
\def\ie{ie.}

\begin{document}

\begin{frontmatter}

\title
  {
  {\bf
  Automatic Classification of Human Epithelial Type 2 Cell\\
  Indirect Immunofluorescence Images using Cell Pyramid Matching
  }
  }

\author
  {
  {Arnold~Wiliem, Conrad~Sanderson, Yongkang~Wong, Peter~Hobson, Rodney~F.~Minchin, Brian~C.~Lovell}\\
  ~\\
  {The University of Queensland, QLD 4072, Australia}\\
  {NICTA, GPO Box 2434, Brisbane, QLD 4001, Australia}\\
  {Queensland University of Technology, QLD 4000, Australia}\\
  {National University of Singapore, Singapore}\\
  {Sullivan Nicolaides Pathology, Australia}
  }

\begin{abstract}

\noindent
This paper describes a novel system for automatic classification of images
obtained from Anti-Nuclear Antibody (ANA) pathology tests on Human Epithelial type 2 (\mbox{HEp-2}) cells using the Indirect Immunofluorescence (IIF) protocol.
The IIF protocol on HEp-2 cells has been the hallmark method to identify the presence of ANAs,
due to its high sensitivity and the large range of antigens that can be detected.
However, it suffers from numerous shortcomings, such as being subjective as well as time and labour intensive.
Computer Aided Diagnostic (CAD) systems have been developed to address these problems,
which automatically classify a \mbox{HEp-2} cell image into one of its known patterns (eg.~speckled, homogeneous).
Most of the existing CAD systems use handpicked features to represent a \mbox{HEp-2} cell image,
which may only work in limited scenarios.
We propose a novel automatic cell image classification method termed Cell Pyramid Matching (CPM),
which is comprised of regional histograms of visual words coupled with the Multiple Kernel Learning framework.
We present a study of several variations of generating histograms
and show the efficacy of the system on two publicly available datasets:
the ICPR \mbox{HEp-2} cell classification contest dataset and the \mbox{SNPHEp-2} dataset.

\end{abstract}

\begin{keyword}
Indirect Immunofluorescence tests \sep Bag of visual words \sep HEp-2 cell classification

\end{keyword}
\end{frontmatter}

\pagestyle{empty}

\section{Introduction}

\begin{figure*}[!tb]
  \begin{minipage}{1\textwidth}
    \centering
    \begin{minipage}{0.23\textwidth}
      \centering
      \includegraphics[width=1.0\textwidth]{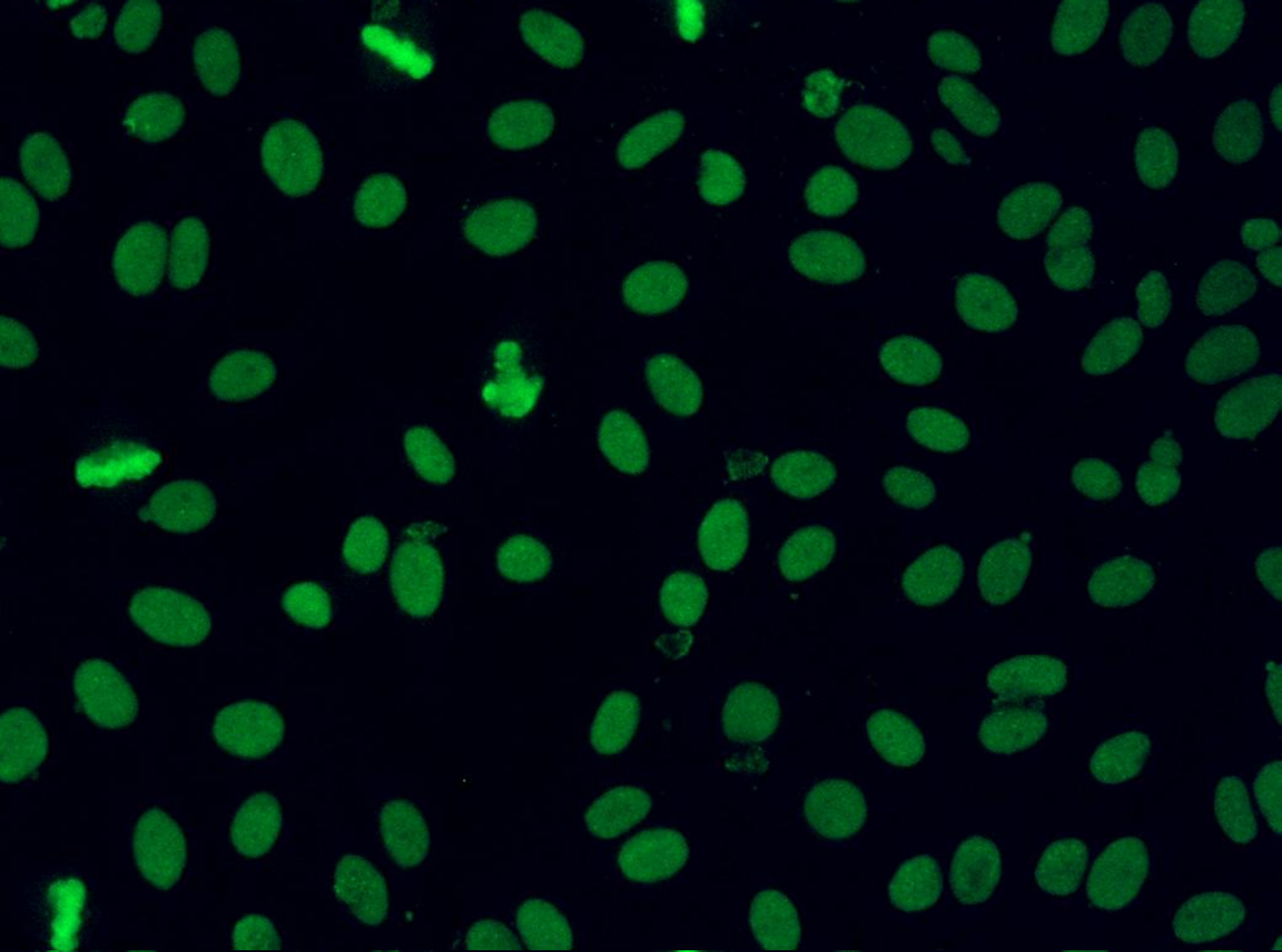}
      {\small homogeneous}
    \end{minipage}    
    \hfill
    \begin{minipage}{0.23\textwidth}
      \centering
      \includegraphics[width=1.0\textwidth]{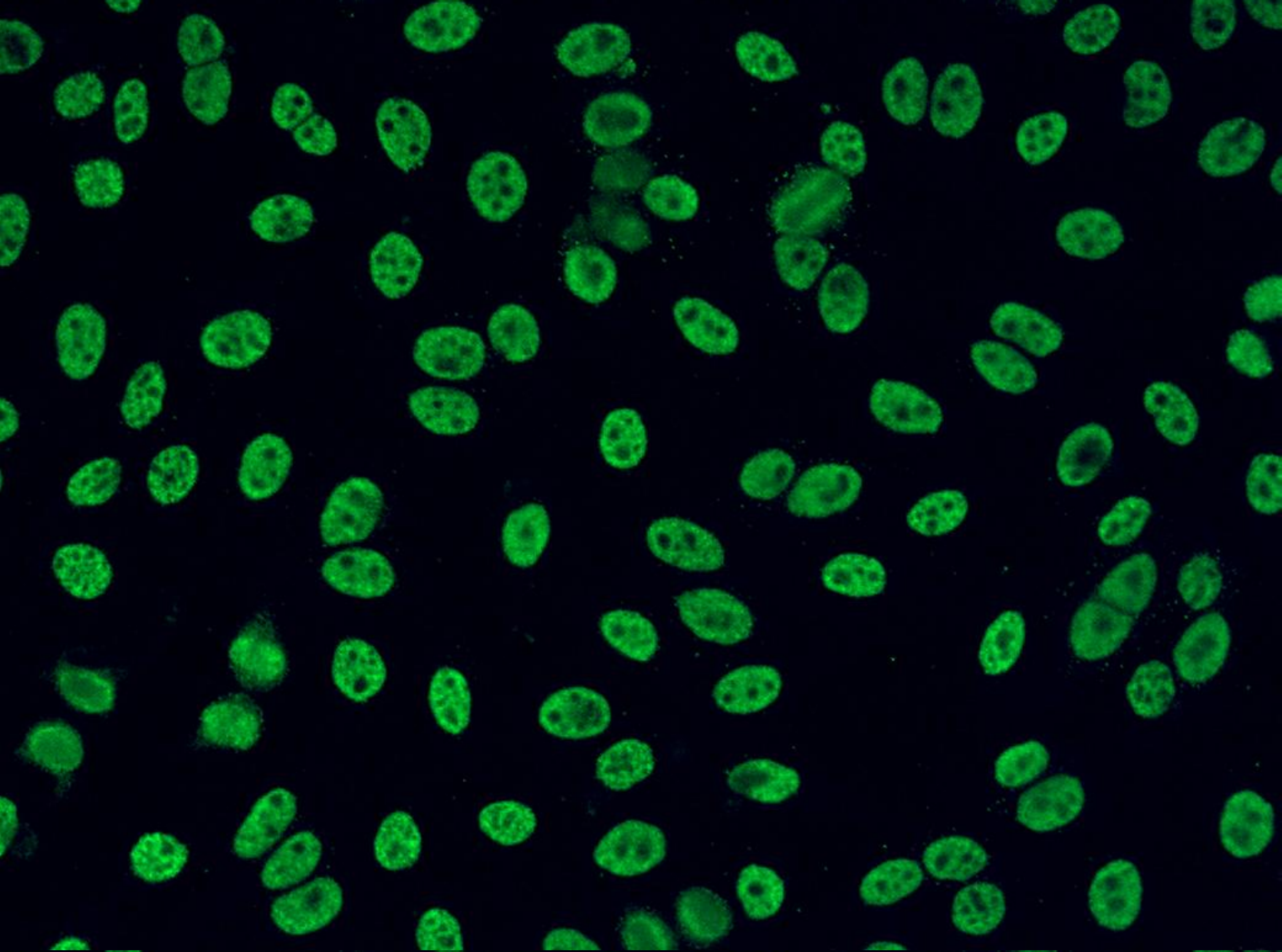}
      {\small speckled}
    \end{minipage}
    \hfill
    \begin{minipage}{0.23\textwidth}
      \centering
      \includegraphics[width=1.0\textwidth]{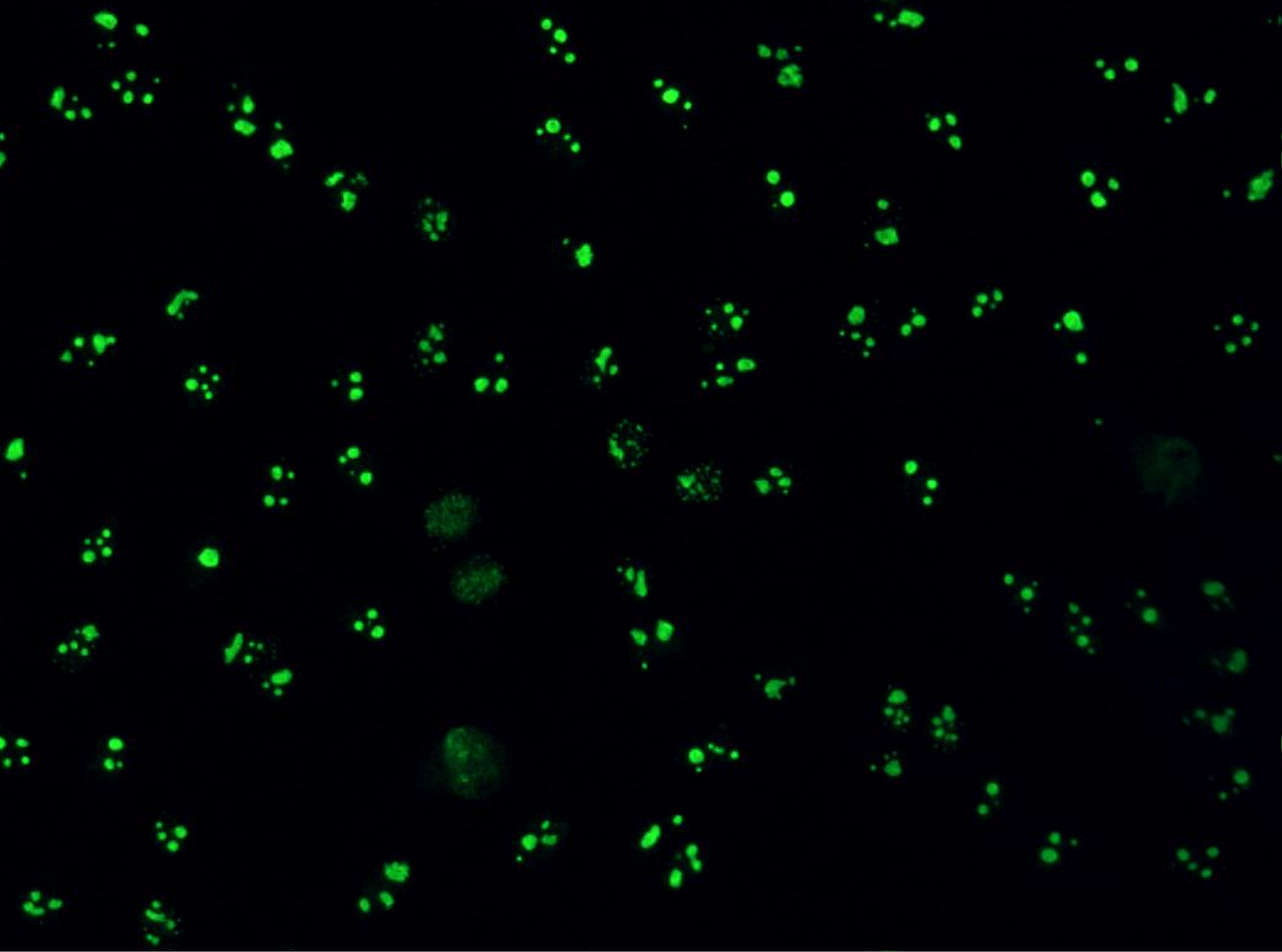}
      {\small nucleolar}
    \end{minipage}
    \hfill
    \begin{minipage}{0.23\textwidth}
      \centering
      \includegraphics[width=1.0\textwidth]{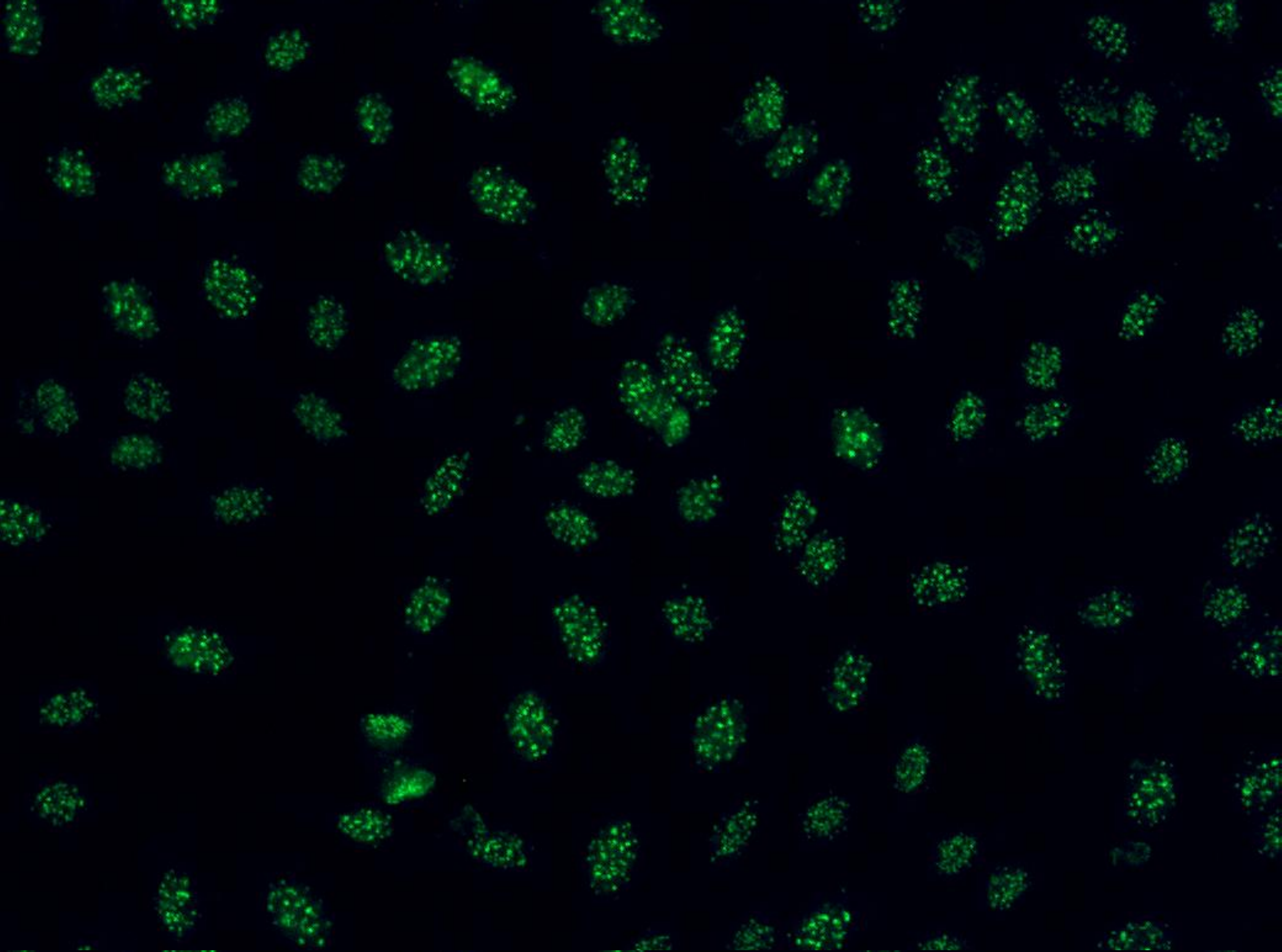}
      {\small centromere}
    \end{minipage}
  \end{minipage}    
  
  \caption{Examples of strong positive ANA specimens. See Fig.~\ref{fig:dataset} for images of individual cells.}  
  \label{fig:specimen_images}
\end{figure*}

The Anti-Nuclear Antibody (ANA) test is commonly used by clinicians to identify the existence of Connective Tissue
Diseases such as Systemic Lupus Erythematosus, Sj\"{o}rgren's syndrome, and Rheumatoid Arthritis~\cite{Meroni2010}.
The hallmark protocol for doing this is through Indirect Immunofluorescence (IIF) on Human Epithelial type 2 (HEp-2) cells~\cite{Meroni2010,Wiik2010}.
This is due to its high sensitivity and the large range expression of antigens.
Examples of specimen images are shown in Figure~\ref{fig:specimen_images}. 
Despite the advantages, the IIF approach is labour intensive and time consuming~\cite{Bizzaro1998,Pham2005}. 
Each ANA specimen must be examined under a fluorescence microscope by at least two scientists. 
This also renders the test result subjective,
and thus has low reproducibility and large variabilities across personnel and laboratories~\cite{Hiemann2009,Soda2009}.

\begin{figure*}[!tb]
  \centering
  \includegraphics[width=0.9\textwidth]{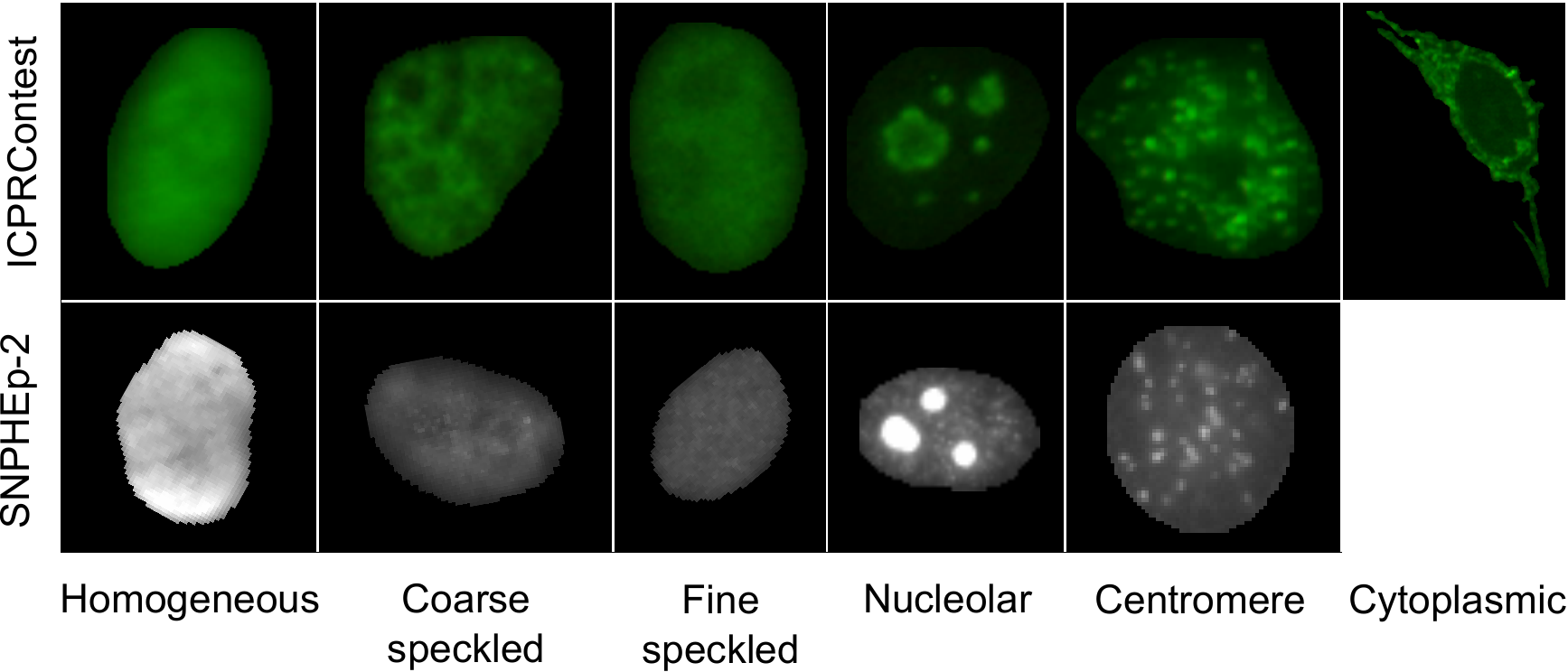}
  \caption{Sample images from ICPRContest dataset~\cite{Foggia2010} and  \mbox{SNPHEp-2} dataset.}
  \label{fig:dataset} 
\end{figure*}

In recent years, there has been increasing interest in employing image analysis 
techniques for various routine clinical pathology tests~\cite{Hiemann2009,Gurcan2009,Khutlang2010}. 
Results produced by these techniques can be used to support the scientists' manual/subjective analysis,
leading to test results being more reliable and consistent across laboratories~\cite{Hiemann2009}. 
Thus, in order to address the shortcomings of the manual test procedure, one could use Computer Aided Diagnostic (CAD) systems which automatically determine the pattern in
the given \mbox{HEp-2} cell images of a specimen~\cite{Hiemann2009,Soda2009,Cordelli2011,Elbischger2009,Hsieh2009,Perner2002a,Strandmark2012,Wiliem2013}.

Table~\ref{tab:relatedworks} presents notable CAD systems proposed in the literature over the last five years.
Most of these systems use carefully handpicked features which may only work in a particular laboratory environment and/or microscope configuration.
To address this, several approaches employ a large number of features and apply an automated feature selection process~\cite{Hiemann2009}. 
Another approach uses Multi Expert Systems to allow the use of a specifically tailored feature set and classifier for each \mbox{HEp-2} cell pattern class~\cite{Soda2009}. 
Nevertheless, the generalisation ability of these systems is still not guaranteed since these systems 
were only evaluated on a dataset with a specific setup.

\begin{table*}
  \footnotesize
  \caption
    {
    Existing CAD systems for \mbox{HEp-2} cell classification.
    }
  \label{tab:relatedworks}
  \vspace{0.5ex}
  \centering
  \begin{tabular}{lll}
    \toprule
    \textbf{Approach}                                   & ~\textbf{Descriptors}                                                 & ~\textbf{Classifier} \\ 
    \hline
    \hline
    Perner~\etal~\cite{Perner2002a}             & ~Textural                                                 & ~Decision Tree \\
    Hiemann~\etal~\cite{Hiemann2009} & ~Structural; textural                              & ~LogisticModel Tree\\  
    Elbischger~\etal~\cite{Elbischger2009}       & ~Image statistics; cell shape; ~textural   & ~Nearest Neighbour (NN) \\        
    \hline
    Hsieh~\etal~\cite{Hsieh2009}                 & ~Image statistics; textural                     & ~Learning Vector Quantisation (LVQ) \\ 
    Soda~\etal~\cite{Soda2009}                   & ~Specific set of features (e.g. textural) for each class & ~Multi Expert System \\ 
     Cordelli~\etal~\cite{Cordelli2011}           & ~Image statistics; textural; morphological  & ~AdaBoost \\      
    \hline
     Strandmark~\etal~\cite{Strandmark2012}                    & ~Morphological; image statistics; textural  & ~Random Forest \\    
    Ali~\etal~\cite{Ali2012} &~Biological-Inspired Descriptor &~Boosted k-NN Classifier \\ 
    Theodorakopoulos~\etal~\cite{Theodorakopoulos2012} &~Morphological and texture features &~Kernel SVM (KSVM) \\
    \hline
    Thibault~\etal~\cite{Thibault2012}&~Morphological and texture features &~Linear Regression, Random Forest\\
    Ghosh~\etal~\cite{Ghosh2012}&~Histograms of Oriented Gradients, &~SVM\\
                              ~ &~image statistics and textural &~\\
    Li~\etal~\cite{Li2012} &~Textural and image statistics &~SVM\\
    \hline
    Di Cataldo~\etal~\cite{DiCataldo2012} &~GLCM and DCT features &~SVM\\ 
    Snell~\etal~\cite{Snell2012} &~Texture and shape &~Multistage classifier\\  
    Ersoy~\etal~\cite{Ersoy2012}&~Local shape measures, gradient and textural &~ShareBoost\\    
    Wiliem~\etal~\cite{Wiliem2013}               &~Bag of visual words with dual-region structure                  &~Nearest Convex Hull Classifier (NCH) \\
    \bottomrule
  \end{tabular}
  \vspace{-2ex}
\end{table*}

One of the most popular approaches for automatic image classification,
here called the bag-of-visual-words (BoW) approach,
is to represent an image in terms of a set of visual words,
selected from a dictionary that has been trained beforehand~\cite{Lazebnik2006,Sanderson2009,Gemert2010,Wong_IJCNN_2012}. 
In order to model an image, the BoW approach divides the image into small image patches, 
followed by patch-level feature extraction. 
An encoding process is then employed to compute a histogram of occurrences of visual words based on these patches. 
BoW descriptors often have higher discrimination power compared to the other image
descriptors~\cite{Wiliem2013,Lazebnik2006,Gemert2010,Wong_IJCNN_2012}.
However, the BoW descriptor has many design options.
For example, one needs to determine which patch-level features and encoding technique is most suitable for the task at hand.
Our previous study presents an extensive evaluation of popular BoW descriptors
in the literature applied to the domain of cell classification~\cite{Wiliem2013}.

\newpage
A single histogram of visual words of an image only describes the visual word statistics and does not retain spatial information
(ie.~where a visual word appears in the image).
Previous studies suggest that location and scale information can provide meaningful discriminative information~\cite{Lazebnik2006, Yang2009}.
For example, the locations of visual words describing a wheel could be used to infer the type of vehicle
(ie.~whether it is a motorcycle, car, or truck).
Spatial Pyramid Matching (SPM) was proposed to exploit this information~\cite{Lazebnik2006}. 
Specifically, each image is processed as a pyramid of levels, with each level containing non-overlapping regions.
The levels differ from each other through an increasing number of regions. 
Each region is divided into small image patches, and an average histogram of visual words is computed for each region. 
The histograms from all regions are then fed into a Support Vector Machine (SVM) classifier~\cite{Shawe-Taylor:2004:KMP}
that uses a specialised kernel.

Our previous work~\cite{Wiliem2013} proposed a Dual-Region (DR) structure within the BoW framework,
specifically designed for cell images.
Each cell image is divided into two regions:
(1)~an inner area enclosing inside the cell; and
(2)~an outer area containing only the cell edge. 
The use of two regions forces the inner and outer cell content to be modelled and compared separately, 
leading to higher recognition accuracies than using only one average region (ie.~single histogram) for each cell image.
An advantage of this approach is that it has lower dimensionality than SPM (ie.~approximately 90\% less),
leading to considerably lower storage requirements.
However, a mixing coefficient which indicates relative region importance needs to be empirically determined.

The work presented in this paper extends our previous study by proposing
a novel approach termed Cell Pyramid Matching (CPM),
which incorporates the positive aspects of the SPM and DR approaches, while omitting their negative aspects.
Furthermore, we show that combining the CPM approach with a learning framework known as Multiple Kernel Learning~\cite{Rakotomamonjy2008}
(where several variants of CPM are employed concurrently)
leads to state-of-the-art performance on the SNPHEp-2 dataset~\cite{Wiliem2013},
and is comparable to the state-of-the-art on the ICPRContest dataset~\cite{Foggia2010}.

We continue this paper as follows. 
We first delineate the \mbox{HEp-2} cell classification task in Section~\ref{sec:sec_problem}.
In Section~\ref{sec:sec_features} we discuss various forms of BoW descriptors and the proposed CPM approach. 
Section~\ref{sec:sec_experiment} is devoted to experiments and discussions, 
followed by the main findings in Section~\ref{sec:sec_conclusions}.

\section{HEp-2 Cell Classification Task}
\label{sec:sec_problem}

Each positive \mbox{HEp-2} cell image
is represented as a three-tuple {\small$(\Mat{I},\Mat{M},\delta)$} which consists of: 
{\bf (i)}~the Fluorescein Isothiocyanate (FITC) image channel {\small$\Mat{I}$}; 
{\bf (ii)}~a binary cell mask image {\small$\Mat{M}$} which can be manually defined, or extracted from the 
(DAPI) image channel~\cite{Hiemann2009}; and 
{\bf (iii)}~the fluorescence intensity {\small$\delta \in \{\text{strong},\text{weak}\}$} which specifies whether the cell is a strong
positive or weak positive.
Strong positive images normally have more defined details, while weak positive images are duller.

Let {\small$\Mat{Y}$} be a probe image {\small$\Mat{Y} = (\Mat{I},\Mat{M},\delta)$}, and $\ell$ be its class label. 
Given a gallery set
{\small $\mathcal{G} = \{ (\Mat{I},\Mat{M},\delta)^{\mathcal{G}}_1, (\Mat{I},\Mat{M},\delta)^{\mathcal{G}}_2, \ldots, (\Mat{I},\Mat{M},\delta)^{\mathcal{G}}_m\}$},
the task of a classifier {\small$\varphi:\Mat{Y} \times \mathcal{G} \mapsto \widehat{\ell}$} is to produce {\small$\widehat{\ell}$},
where ideally {\small$\widehat{\ell} = \ell$}. 

We consider six \mbox{HEp-2} cell patterns~\cite{Wiik2010} listed below;
example images are shown in Fig.~\ref{fig:dataset}.

\begin{enumerate}[{\bf (1)}]
\renewcommand{\itemsep}{0ex}

\item
{\it homogeneous}: a uniform diffuse fluorescence covering the entire nucleoplasm sometimes accentuated in the nuclear periphery

\item
{\it coarse speckled}: densely distributed, variously sized speckles, generally associated with larger speckles, throughout nucleoplasm of interphase cells;
 nucleoli are negative

\item
{\it fine speckled}: fine speckled staining in a uniform distribution, sometimes very dense so that an almost homogeneous pattern is attained; 
 nucloli may be positive or negative

\item
{\it nucleolar}: brightly clustered larger granules corresponding to decoration of the fibrillar centers of the nucleoli as well as the coiled bodies

\item
{\it centromere}: rather uniform discrete speckles located throughout the entire nucleus

\item
{\it cytoplasmic}: a very fine dense granular to homogeneous staining or cloudy pattern covering part or the whole cytoplasm
\end{enumerate}

\section{Bag of Words Classification Systems}
\label{sec:sec_features}

\begin{figure*}[!tb]
  \centering
  \includegraphics[width=0.7\textwidth]{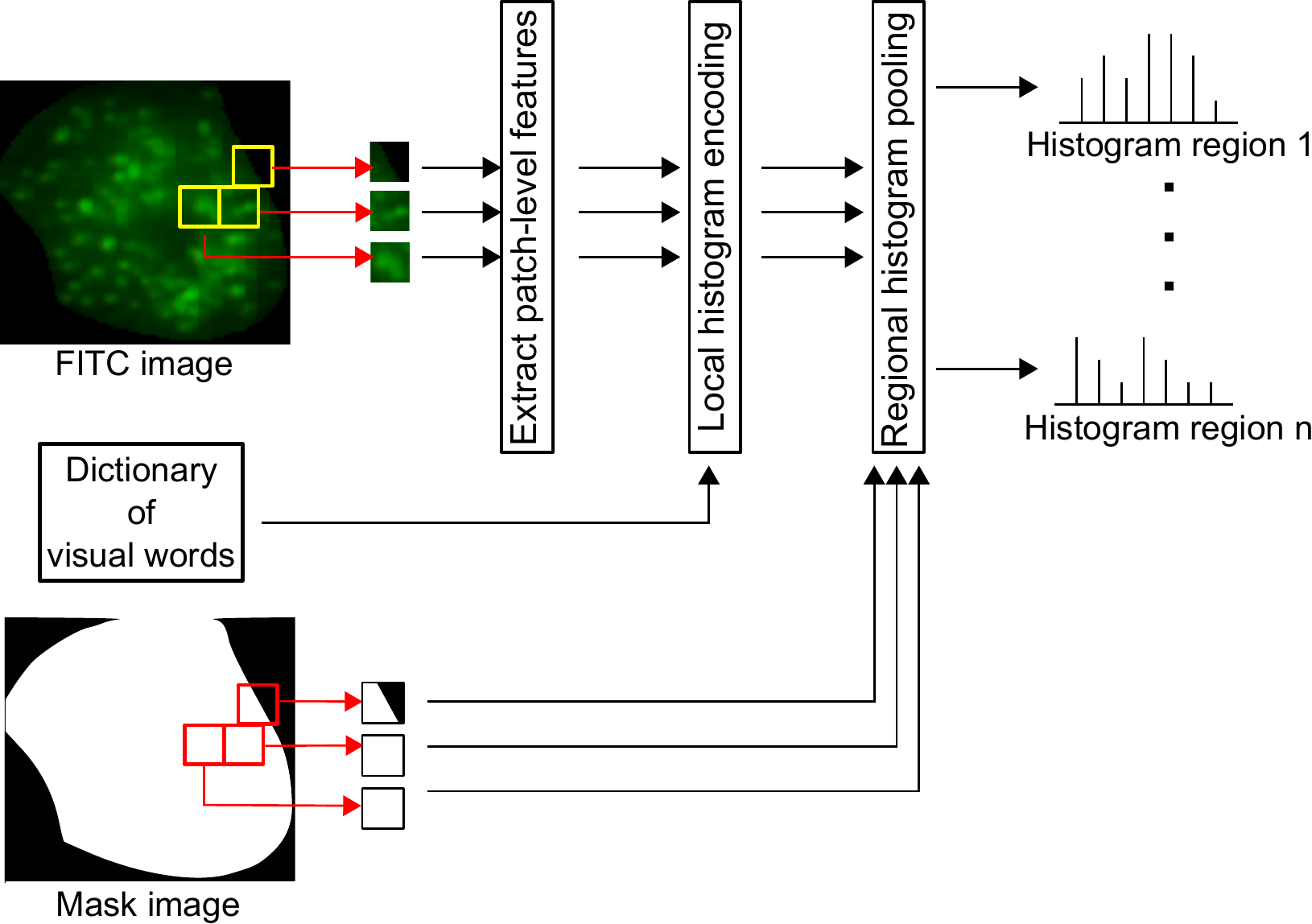} 
  \caption
    {
    Conceptual diagram of the general approach for obtaining histograms of visual words from cell images.
    Both the FITC image and its corresponding mask image are divided into small overlapping patches.  
    Patch-level features are extracted from FITC patches. 
    Local histogram from each FITC patch-level features is obtained by an encoder employing a learned dictionary of visual words.   
    Finally, multiple regional descriptors are then computed by pooling the local histograms of FITC patches belonging to each region.      
    } 
    \label{fig:system_diagram}  
\end{figure*}

\begin{figure*}[!tb]
  \centering
  \includegraphics[width=1\textwidth]{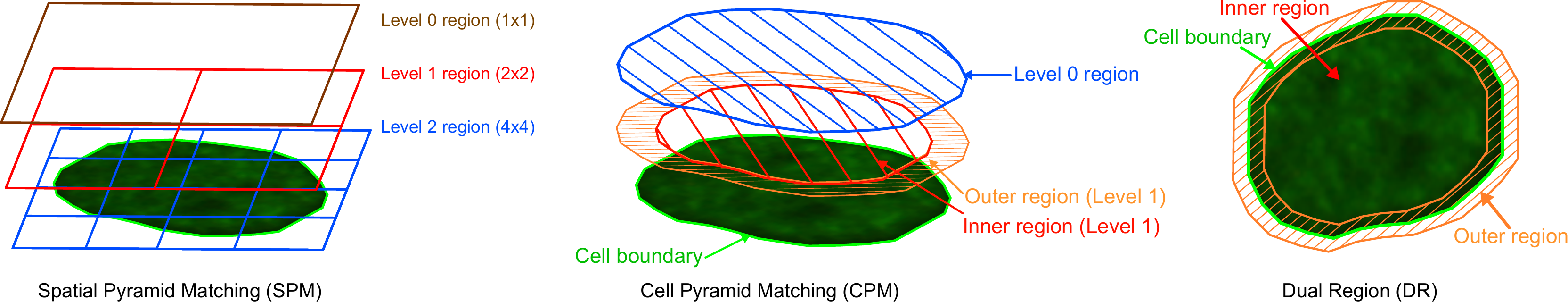}  
  \caption{
  Conceptual diagrams for various spatial structures to obtain multiple region descriptors.  
  }
  \label{fig:regional_descriptor} 
\end{figure*}

A conceptual illustration of the general approach for obtaining histograms of visual words from HEp-2 cell images is shown in Fig.~\ref{fig:system_diagram}.
Each cell image is first resized into a canonical size and then divided into small overlapping patches.
The patches are in turn represented by patch-level features.
The {\it local} histogram from each patch is then extracted by using the pre-trained visual word dictionary.
The local histograms located inside a region are pooled to compute the overall histogram for the region. 
Finally, the cell image is represented by a set of regional histograms;
examples of regional structures are shown in Fig.~\ref{fig:regional_descriptor}.

In the following sub-sections, we first describe low-level patch-level features,
followed by presenting various methods for local histogram extraction.
The regional structures (ie.~SPM, DR and the proposed CPM) are discussed afterwards.
Finally, we overview a framework known as Multiple Kernel Learning (MKL),
which combines information captured from several descriptors. 

\subsection{Patch-level Feature Extraction}

Given a HEp-2 cell image {$(\Mat{I},\Mat{M},\delta)$},
both the FITC image $\Mat{I}$ and mask image $\Mat{M}$ are divided into small overlapping patches
{$\mathcal{P}_I = \{\Mat{p}_{I,1},~ \Mat{p}_{I,2},~ \ldots,~ \Mat{p}_{I,n}\}$}
and
{$\mathcal{P}_M = \{\Mat{p}_{M,1},~ \Mat{p}_{M,2},~ \ldots,~ \Mat{p}_{M,n}\}$}.
The division is accomplished in the same manner of both images,
resulting in each patch in the FITC image having a corresponding patch in the mask image.
Let {$\Vec{f}$} be a patch-level feature extraction function
{$\Vec{f}: \Mat{p}_{I} \mapsto \Vec{x}$}, where {$\Vec{x} \in \mathbb{R}^d$}. 
{$\mathcal{P}_{I}$} now can be represented as {$\mathcal{X}=\{\Vec{x}_1,~ \Vec{x}_2,~ \ldots,~ \Vec{x}_n\}$}.

For evaluation purposes, we selected two popular patch-level feature extraction techniques,
based on the Scale Invariant Feature Transform (SIFT)
and the Discrete Cosine Transform (DCT). 
The SIFT descriptor is invariant to uniform scaling, orientation
and partially invariant to affine distortion and illumination changes~\cite{Lowe2004a}.
These attributes are advantageous in this classification task as cell images are unaligned and have high within class variabilities. 
DCT based features proved to be effective for face recognition in video surveillance~\cite{Sanderson2009,Wong_IJCNN_2012}.
By using only the low frequency DCT coefficients (essentially a low-pass filter),
each patch representation is relatively robust to small alterations~\cite{Sanderson2009}. 
We follow the extraction procedures for SIFT and DCT as per~\cite{Liu2011} and~\cite{Sanderson2009}, respectively.

The dictionary of visual words, denoted as $\mathcal{D}$,
is trained from patches extracted in sliding window manner from training cell images.
Each histogram encoding method has specific dictionary training procedure.

\subsection{Generation of Local Histograms}
\label{sec:histogram_generation}

For each patch-level feature that belongs to region $r$, {$\Vec{x}_j \in \mathcal{X}_r$},
a local histogram {$\Vec{h}_j$} is obtained.
In this work we consider three prominent histogram encoding methods:
(1)~vector quantisation; (2)~soft assignment; (3)~sparse coding. 
The methods are elucidated below.

\subsubsection{Vector Quantisation (VQ)}

Given a set {$\mathcal{D}$}, the dictionary of visual words,
the $i$-th dimension of local histogram {$\Vec{h}_j$} for patch {$\Vec{x}_j$} is computed via:

\vspace{-1ex}
\begin{small}
\begin{equation}
  \Vec{h}_{j,i} = \left\{
  \begin{array}{l l}
  1 & \quad \text{if } i = \argmin{k \in 1, \ldots, |\mathcal{D}|}  { \operatorname{dist}(\Vec{x}_j,\Vec{d}_k)} \\
  0 & \quad \text{otherwise}
  \end{array}
  \right.
  \label{eqn:vq_encoding}
\end{equation}
\end{small}

\vspace{-1ex}
\noindent 
where {$\operatorname{dist}(\Vec{x}_j,\Vec{d}_k)$}
is a distance function between {$\Vec{x}_j$} and {$\Vec{d}_k$},
while {$\Vec{d}_k$} is the {$k$}-th entry in the dictionary {$\mathcal{D}$} 
and {$|\mathcal{D}|$ is the number of elements in {$\mathcal{D}$}.
The dictionary is obtained via the $k$-means algorithm~\cite{Bishop2006} on training patches,
with the resulting cluster centers representing the entries in the dictionary. 

The VQ approach is considered as a hard assignment approach since each image patch is only assigned to one of the visual words.
Such hard assignment can be sensitive to noise~\cite{Gemert2010}. 

\subsubsection{Soft Assignment (SA)}
\label{sec:soft_assignment}

In comparison to the VQ approach above,
a more robust approach is to apply a probabilistic method~\cite{Sanderson2009}.
Here the visual dictionary {$\mathcal{D}$} is a convex mixture of Gaussians.
The $i$-th dimension of the local histogram for {$\Vec{x}_j$} is calculated by: 

\vspace{-1ex}
\begin{small}
\begin{equation}
  \Vec{h}_{j,i} = \frac{w_i p_i(\Vec{x}_j)}{\sum_{k=1}^{|\mathcal{D}|}{ w_k ~ p_k( \Vec{x}_j ) }}
  \label{eqn:probabilistic_encoding}
\end{equation}
\end{small}

\vspace{-1ex}
\noindent
where {$p_i( \Vec{x} )$} is the likelihood of {$\Vec{x}$} according to the $i$-th component of the visual dictionary $\mathcal{D}$:

\vspace{-1ex}
\begin{small}
\begin{equation}
  p_i(\Vec{x})  = \frac
    {
    \exp \left[ -\frac{1}{2}\left(  \Vec{x} - \Vec{\mu}_i \right)^T \Mat{C}_i^{-1} \left(
    \Vec{x} - \Vec{\mu}_i \right) \right] }
    {
    \left( 2 \pi \right)^\frac{d}{2} | \Mat{C}_i | ^\frac{1}{2}
    }
\end{equation}
\end{small}

\vspace{-1ex}
\noindent
with {$w_i$}, {$\Vec{\mu}_i$} and {$\Mat{C}_i$}
representing the weight, mean and diagonal covariance matrix of Gaussian $i$, respectively.
The scalar $d$ represents the dimensionality of {$\Vec{x}$}.
The dictionary {$\mathcal{D}$} is obtained using the Expectation Maximisation algorithm~\cite{Bishop2006} on training patches.

\subsubsection{Sparse Coding (SC)}

It has been observed that each local histogram produced via Eqn.~(\ref{eqn:probabilistic_encoding})
is sparse in nature (ie.~most elements are close to zero)~\cite{Wong_IJCNN_2012}.
In other words, the SA approach described in Section~\ref{sec:soft_assignment} is an {\it indirect} sparse coding approach.
Hence, it is possible to adapt \textit{direct} sparse coding algorithms
in order to represent each patch as a combination of dictionary atoms~\cite{Coates2011,Yang2009},
which theoretically can lead to better recognition results~\cite{Wong_IJCNN_2012}.

A vector of weights {$\Vec{\vartheta} = \left[ \vartheta_1, \vartheta_2,...,\vartheta_{|\mathcal{D}|} \right]^T$}
is computed for each {$\Vec{x}_j$} by solving a minimisation problem that selects a sparse set of dictionary atoms. 
As the theoretical optimality of the \mbox{$\ell_1$-norm} minimisation solution is guaranteed~\cite{Tropp2010},
in this work we used:

\vspace{-1ex}
\begin{small}
\begin{equation}
  \operatorname{min}\frac{1}{2}\|\Mat{D}\Vec{\vartheta}-\Vec{x}_j\|^2_2 + \lambda \sum\nolimits_k{\|\vartheta_k\|_1}      
  \label{eq:SCObjective}
\end{equation}
\end{small}

\vspace{-1ex}
\noindent
where
{$\| \cdot \|_p$} denotes the \mbox{$\ell_p$-norm}
and
{$\Mat{D} \in \mathbb{R}^{d \times |\mathcal{D}|}$} is a matrix of dictionary atoms.
The dictionary {$\Mat{D}$} is trained by using the K-SVD algorithm~\cite{Aharon2006},
which is known to be suitable for obtaining reasonable dictionaries in similar cases,
ie.,~using a large number of small image patches~\cite{Rubinstein2010}.

As $\Vec{\vartheta}$ can have negative values due to the objective function in Eqn.~(\ref{eq:SCObjective}),
we construct each local histogram using the absolute value of each element in {$\Vec{\vartheta}$}~\cite{Wong_IJCNN_2012}:

\vspace{-1ex}
\begin{small}
\begin{equation}
  \Vec{h}_j = \left[~ |\vartheta_1|,~ |\vartheta_2|,~ \ldots,~ |\vartheta_{|\mathcal{D}|}|~ \right]
\end{equation}
\end{small}

\vspace{-1ex}
Compared to both Eqns.~(\ref{eqn:vq_encoding}) and~(\ref{eqn:probabilistic_encoding}),
obtaining the histogram using sparse coding is considerably more computationally intensive,
due to the need to solve a minimisation problem for each patch.

\subsection{Histogram Pooling}

Let {$\mathcal{X}_r$} be the set of patch-level features belonging to region~$r$.
The overall histogram representation for region $r$ is then obtained via
averaging local histograms~\cite{Sanderson2009,Wong_IJCNN_2012}:

\vspace{-1ex}
\noindent
\begin{small}
\begin{equation}
  \Vec{H}^{[r]} = \frac{1}{|\mathcal{X}_r|} \sum\nolimits_{j=1}^{|\mathcal{X}_r|} \Vec{h}_j
\end{equation}
\end{small}

\vspace{-1ex}
\noindent
where {$|\mathcal{X}_r|$} is the number of elements in set {$\mathcal{X}_r$}.
In the following subsections, we describe several possible spatial layouts for the regions
and the corresponding similarity measures.

\subsection{Spatial Structures for Multiple Region Descriptors}
\label{sec:region_structure}

In this section we describe two existing spatial structures for using multiple regional descriptors (ie.~SPM and DR),
followed by the proposed CPM approach.
The conceptual diagram for each approach is shown in Fig.~\ref{fig:regional_descriptor}.

\subsubsection{Spatial Pyramid Matching (SPM)}

The regions are organised similar to an image pyramid with several levels~\cite{Lazebnik2006}. 
At each level $l$, the image is divided into $(2^l) \times (2^l)$ non-overlapping regions.
For instance, at level 0 (ie.~the top level), the image is divided into 1$\times$1 region;
at level 1, the image is divided into 2$\times$2 regions. 
In this work, we follow Lazebnik~\etal~\cite{Lazebnik2006} by using a three-level pyramid (\ie~levels 0, 1 and 2):
1$\times$1, 2$\times$2 and 4$\times$4.
In total, there are $1 + 4 + 16 = 21$ regions. 
The {\it pyramid match kernel} is used to measure the similarities between two images~\cite{Lazebnik2006}:

\noindent
\begin{small}
\begin{equation}
\mathcal{K}(\Mat{H}_1,\Mat{H}_2)
=
\frac{1}{2^L} G \left( \Vec{H}^{[0,r]}_{1}, \Vec{H}^{[0,r]}_{2} \right)
+
\sum^L_{l=1}{\frac{1}{2^{L-l+1}} G \left( \Vec{H}^{[l,r]}_{1}, \Vec{H}^{[l,r]}_{2} \right) }
\label{eq:SPMKernel}
\end{equation}%
\end{small}%

\noindent
where $\Mat{H}^{[l,r]}_{k}$ is the $r$-th regional histogram of levels $l$ of the $k$-th image,
while $L$ is the maximum number of levels (\ie~$L = 2$). 
$G(\cdot,\cdot)$, is a histogram intersection kernel, defined as~\cite{Lazebnik2006}:

\noindent
\begin{small}
\begin{equation}
G \left( \Mat{H}^{[l,r]}_{1},\Mat{H}^{[l,r]}_{2} \right) = \sum_{j} \operatorname{min} \left( \Vec{H}^{[l,r]}_{1,j}, \Vec{H}^{[l,r]}_{2,j} \right)  
\end{equation}%
\end{small}%

\noindent
where $\Vec{H}^{[l,r]}_{k,j}$ is $j$-th dimension of a regional histogram for level $l$ and region $r$ of image $k$.

\subsubsection{Dual Region (DR)}
Each cell is divided into an inner region, which covers the cell content,
and an outer region, which contains information related to cell edges and shape~\cite{Wiliem2013}.
To this end, each patch is first classified as either belonging to the inner or outer region by inspecting its corresponding mask patch.
More specifically, let {$\mathcal{X} = \mathcal{X}^{[o]} \cup \mathcal{X}^{[i]}$},
with {$\mathcal{X}^{[o]}$} representing the set of outer patches,
and {$\mathcal{X}^{[i]}$} the set of inner patches.
The classification of patch {$\Mat{p}$} into a region is done via:

\noindent
\begin{small}
\begin{equation}
  \Mat{p}_{I} \in  \left\{
  \begin{array}{l l}  
  \mathcal{X}^{[o]} & \quad \text{if} ~ \operatorname{fg}(\Mat{p}_{M}) \in [\tau_1, \tau_2) \\ 
  \mathcal{X}^{[i]} & \quad \text{if} ~ \operatorname{fg}(\Mat{p}_{M}) \in [\tau_2, 1]\\  
  \end{array}
  \right.  
  \label{eqn:patch_classification}
\end{equation}%
\end{small}%

\noindent
where {$\Mat{p}_{M}$} is the corresponding mask patch;
{$\operatorname{fg}(\Mat{p}_{M}) \in [0,1]$} computes the normalised occupation count of foreground pixels from mask patch {$\Mat{p}_{M}$};
$\tau_1$ is the minimum foreground pixel occupation of a patch belonging to the outer region;
$\tau_2$ is the minimum pixel occupation of a patch belonging to the inner region.
Note that the size of the inner and outer regions is indirectly determined via Eqn.~(\ref{eqn:patch_classification}).
Based on preliminary experiments, we have found that $\tau_1 = 0.3$ and $\tau_2 = 0.8$ provide good results.
Unlike SPM, there are only two regional histograms required to represent a cell image. 
As such, the DR descriptor is $(21-2)/21\approx90\%$ smaller than SPM.

The similarity between two images is defined via:

\noindent
\begin{small}
\begin{equation}
  \mathcal{K}(\Mat{H}_1,\Mat{H}_2) = \operatorname{exp}\left[-\operatorname{dist}\left(\Mat{H}_1,\Mat{H}_2\right) \right]
  \label{eq:MCKKernel}
\end{equation}%
\end{small}%

\noindent
Adapting~\cite{Sanderson2009}, $\operatorname{dist}(\Mat{H}_1,\Mat{H}_2)$ is defined by:

\noindent
\begin{equation}
  \operatorname{dist}(\Mat{H}_1,\Mat{H}_2) = \alpha^{[i]} \| \Vec{H}^{[i]}_{1} - \Vec{H}^{[i]}_{2} \|_1 + \alpha^{[o]} \| \Vec{H}^{[o]}_{1} - \Vec{H}^{[o]}_{2} \|_1
\end{equation}%

\noindent
where $\Vec{H}^{[i]}_{k}$ and $\Vec{H}^{[o]}_{k}$ are the inner and outer region histograms of image $k$, respectively; 
$\alpha^{[i]}$ and $\alpha^{[o]}$ are positive mixing parameters which define the importance of information contained for each region,
under the constraint of $\alpha^{[i]} + \alpha^{[o]} = 1$.
A possible drawback of the DR approach is that determining good settings for the $\tau_1$, $\tau_2$ and $\alpha^{[i]}$ parameters 
is currently a time consuming procedure, where a heuristic or grid-based search is used.
Furthermore, not all valid settings in such a search might be evaluated, which can lead to sub-optimal discrimination performance.

\subsubsection{Cell Pyramid Matching (CPM)}
\label{sec:cell_pyramid_matching}

The proposed CPM approach combines the advantages of both SPM and DR structures. 
It adapts the idea of using a pyramid structure from SPM as well as the inner and outer regions from DR.
Unlike SPM, CPM has only two levels: level 0 which comprises the whole cell region,
and level 1 which comprises of inner and outer regions.
The advantages of this combination are two fold:
(1)~the CPM descriptor only requires 3 histograms to represent a cell image,
and is hence $(21-3)/21\approx85\%$ smaller than SPM;
(2)~as the CPM follows the SPM construct, it employs the pyramid match kernel,
which eliminates the mixing parameters in DR.

\subsection{Multiple Kernel Learning}
\label{sec:MKL}

Fusing information provided by various image descriptors and spatial structures
(each with a dedicated kernel, as shown above) may improve discrimination ability,
if the given descriptors are at least partially capturing differing information.
To that end we have elected to use the Multiple Kernel Learning (MKL) framework,
which aims to learn the optimum mixing of various kernels~\cite{Rakotomamonjy2008}.
Let $\{\Vec{x_i},\Vec{y_i}\} \in \mathcal{G}$ be the training set,
where $\Vec{x_i}$ is a feature vector and $\Vec{y_i} \in \{-1,+1\}$ is the corresponding groundtruth label%
\footnote
  {
  Here we have presented a binary classification problem.
  However, it can be easily generalised into a multi-class problem~\cite{Rakotomamonjy2008}.
  }%
.

The MKL classifier is an extended form of the SVM classifier, defined as:

\noindent
\begin{small}
\begin{equation}
  \operatorname{\varphi}(\Vec{q}) = \sum\nolimits^n_{k=1}{\beta_k\mathcal{K}(\Vec{q},\Vec{x_k})} + \Vec{b} 
  \label{eqn:MKL_1}
\end{equation}%
\end{small}%

\noindent
where $\Vec{q}$ is a query point,
$\Vec{x_k} \in \mathcal{G}$ is the $k$-th training point,
$\beta_k$ is the ``importance'' weight of the $k$-th training point,
$\Vec{b}$ is the bias term,
and $\mathcal{K}(\cdot,\cdot)$ is a combination kernel defined as:

\noindent
\begin{small}
\begin{equation}
  \mathcal{K}(\Vec{a},\Vec{c}) = \sum\nolimits^M_{m=1}{w_m\mathcal{K}_m(\Vec{a},\Vec{c})}
  \label{eqn:MKL_2}
\end{equation}%
\end{small}%

\noindent
where $\mathcal{K}_m(\cdot,\cdot)$ is the $m$-th kernel,
with $w_m$ its corresponding mixing weight,
under the constraints of $w_m \geq 0$ and $\sum{w_m} = 1$.
Without losing generality,
$\mathcal{K}_m(\cdot,\cdot)$ can be the kernel defined in Eqn.~(\ref{eq:SPMKernel}) or~(\ref{eq:MCKKernel}).

In the MKL learning scheme, the importance weights and kernel mixing weights are learned together. 
In this work we employ the SimpleMKL method for learning~\cite{Rakotomamonjy2008},
which employs a convex and smooth objective function.
\section{Experiments and Results}
\label{sec:sec_experiment}

In this section we first compare the performance of six variants of the BoW descriptor,
where each of the two low-level feature extraction techniques (SIFT and DCT)
is coupled with three possible methods for generating the histograms of visual words (VQ, SA, and SC).
The six variants are used within the framework of the DR, SPM and CPM spatial structures.
We then show that by fusing the two approaches (DCT-SA CPM and DCT-VQ CPM)
via the MKL framework leads to an increase in recognition rates.
Finally, we compare the MKL based system against three recently proposed systems in the literature.
The various systems were implemented with the aid of the Armadillo C++ library~\cite{Armadillo_2010}.

\subsection{Datasets: ICPRContest and SNP \mbox{HEp-2}}

For the experiments we used two publicly available datasets, briefly described below,
in order to evaluate applicability of the various systems to differing assays and microscope parameters.

The ICPR \mbox{HEp-2} Cell Classification Contest (ICPRContest) Dataset~\cite{Foggia2013}
contains 1,457 cells extracted from 28 specimen images\footnote
  {
  It is assumed that the cell images have been extracted from specimen images 
  either via a manual or automated approach such as background subtraction~\cite{Reddy_AVSS_2010,Reddy_TCSVT_2012}.
  }.
It contains six patterns: centromere, coarse speckled, cytoplasmic, fine speckled, homogeneous, and nucleolar. 
Each specimen image was acquired by means of fluorescence microscope (40-fold magnification) coupled with 50W mercury vapour lamp and with a CCD camera. 
The cell image masks were hand labelled. 
See Fig.~\ref{fig:dataset} for examples. 
We followed the ICPR contest evaluation protocol for this dataset which only has one pair of train and test sets.

The SNP \mbox{HEp-2} Cell (SNPHEp-2) Dataset%
\footnote{The SNPHEp-2 dataset is available for download at \href{http://staff.itee.uq.edu.au/lovell/snphep2/}{\tt http://staff.itee.uq.edu.au/lovell/snphep2/}}~\cite{Wiliem2013} 
was obtained between January and February 2012 at Sullivan Nicolaides Pathology laboratory, Australia. 
This dataset
has five patterns: centromere, coarse speckled, fine speckled, homogeneous and nucleolar. 
The 18-well slide of HEP-2000 IIF assay from Immuno Concepts N.A.~Ltd.~with screening dilution 1:80 was used to prepare 40 specimens. 
Each specimen image was captured using a monochrome high dynamic range cooled microscopy camera,
which was fitted on a microscope with a plan-Apochromat 20x/0.8 objective lens and an LED illumination source. 
\mbox{\it 4',6-diamidino-2-phenylindole} (DAPI) image channel was used to automatically extract the cell image masks. 

There are 1,884 cell images extracted from 40 specimen images. 
The specimen images are divided into training and testing sets with 20 images each (4 images for each pattern). 
In total there are 905 and 979 cell images extracted for training and testing.
Five-fold validations of training and testing were created by randomly selecting the training and test images. 
Both training and testing in each fold contain around 900 cell images (approx. 450 cell images each).
Examples are shown in Fig.~\ref{fig:dataset}. 

Due to possible varying filtering effects caused by image capture equipment, tuning, operator bias, and/or environmental conditions
(all of which can result in low-pass filtering),
cell images with the same pattern can simply differ due to gross mismatches in frequency spectra.
In turn this can lead to a degradation in recognition accuracy~\cite{Wong_ICPR_2010}.
To counteract this undesirable effect, and to ensure a canonical image size is used,
images from both datasets were downsampled by two to approximately $64 \times 64$ pixels.

\begin{table*}[!tb]
  \centering  
  \caption
    {
    Performance comparison of BoW descriptor variants on the ICPRContest and SNPHEp-2 datasets,
    using various spatial configurations (DR, SPM, CPM).
    The scores for SNPHEp-2  dataset shown as average correct classification rate.
    DR = dual region; SPM = Spatial Matching Pyramid; CPM = Cell Pyramid Matching.
    }
  \label{tab:spatial_structure}
  \vspace{0.5ex}
  \begin{tabular}{l|ccc|cccc}
    \toprule
    \textbf{Descriptor} & \multicolumn{3}{c|}{\textbf{ICPRContest}} & \multicolumn{3}{c}{\textbf{SNPHEp-2}} \\               
    \textbf{Variant}    & DR & SPM & CPM  & DR & SPM & CPM  \\ 
    \midrule
    DCT-SA   & 64.9 & 64.3 & \textbf{65.9} & 79.5 & 80.3 & \textbf{81.2}  \\ 
    DCT-VQ   & 54.5 & 57.1 & \textbf{61.2} & 80.7 & 77.9 & \textbf{80.8}  \\ 
    DCT-SC   & 52.6 & \textbf{57.9} & 57.2 & 71.0 & 70.5 & \textbf{73.5}  \\ 
    \midrule 
    SIFT-SA  & 51.6 & \textbf{57.5} & 47.8 & 71.6 & 69.7 & \textbf{73.2} \\ 
    SIFT-VQ  & 55.6 & 53.8 & \textbf{59.0} & 64.9 & 74.4 & \textbf{75.0} \\ 
    SIFT-SC  & 60.8 & 59.9 & \textbf{62.1} & 76.2 & 73.6 & \textbf{76.3} \\ 
    \bottomrule
  \end{tabular}
\end{table*}

\subsection{Combinations of Local Features, Histogram Generation and Spatial Structures}
\label{sec:spatial_structure_variants}

We follow Lazebnik~\etal~\cite{Lazebnik2006} and Wiliem~\etal~\cite{Wiliem2013} for SPM and DR implementations, respectively.
The SVM classifier is used in all cases, with the kernels specified in Eqns.~(\ref{eq:SPMKernel}) and~(\ref{eq:MCKKernel})
for the SPM and DR methods, respectively.
As noted in Section~\ref{sec:cell_pyramid_matching}, a form of Eqn.~(\ref{eq:SPMKernel}) is used as the SVM kernel for the CPM method.

As there are three histogram encoding methods
(ie.~VQ, SA and SC) and two patch-level features (\ie~SIFT and DCT),
there are six variants of the BoW descriptor. 
For clarity, each variant is denoted by:
\textit{[patch-level features]}-\textit{[histogram encoding method]}.
For example, the variant using DCT as its patch-level features and VQ as its encoding method is called DCT-VQ.

The results, presented in Table~\ref{tab:spatial_structure}, 
indicate that in most cases the proposed CPM system obtains the best performance,
suggesting that it is taking advantage of both the specialised spatial layout for cells inherited from the DR approach,
and the pyramid match kernel inherited from the SPM approach.
The results also show that in most cases the use of DCT based patch-level feature extraction
leads to better performance than using SIFT based feature extraction.
We conjecture that DCT obtains better performance as the SIFT descriptor needs a larger spatial support
and is hence more likely to be affected by image deformations.
Specifically, SIFT divides a given image patch into 4$\times$4 subregions in which each has 4$\times$4 bins,
followed by extracting gradient information from each subregion~\cite{Lowe2004a}. 
Therefore, SIFT needs a spatial support of at least 16$\times$16 pixels,
which is relatively large when compared to the canonical cell image size of 64$\times$64.
In contrast, standard DCT requires a much smaller spatial support of 8$\times$8 pixels,
making it less susceptible to image deformations.

\subsection{Fusion via Multiple Kernel Learning}

Based on the results obtained in the previous experiment,
we have selected the overall top three systems (DCT-SA CPM, DCT-VQ CPM, SIFT-SC CPM)
and evaluated fusing them via the MKL framework.
The results for various mixtures of the three systems are shown in Table~\ref{tab:mkl_combos}.

By using the mixture that obtains the best overall performance across both datasets,
ie.~DCT-SA CPM and DCT-VQ CPM, 
the recognition rate improves from 65.9\% to 67.4\% on the ICPRContest dataset,
and from 81.2\% to 82.4\% on the SNPHEp-2 dataset.

Note that while it is possible to fuse information from more systems,
there is no guarantee that this will always lead to better performance~\cite{Gehler2009,Rakotomamonjy2008,Sanderson_DSP_2004}.
In further experiments (not shown here) we have found that combining more systems
can decrease performance.
As the main aim was to show the possible advantage of using MKL, we leave a detailed study for future work.

\begin{table*}
  \centering
  \caption
    {
    Performance of various systems fused via the MKL framework
    on the ICPRContest and SNPHEp-2 datasets.
    \mbox{The ``overall'' column is the mean performance across the two datasets.}
    }
  \label{tab:mkl_combos}
  \vspace{0.5ex}
  \begin{tabular}{l|c|c|c}
  \toprule
  {\bf System Mixture}                        & {\bf ICPRContest} & {\bf SNPHEp-2} & {\bf overall}  \\
  \midrule
  (DCT-SA CPM) + (DCT-VQ CPM) + (SIFT-SC CPM) & 66.9              & 82.5           & 74.70       \\
  (DCT-SA CPM) + (DCT-VQ CPM)                 & 67.4              & 82.4           & 74.90       \\
  (DCT-SA CPM) + (SIFT-SC CPM)                & 66.3              & 81.2           & 73.75       \\
  (DCT-VQ CPM) + (SIFT-SC CPM)                & 64.0              & 79.7           & 71.85       \\
  \hline
  \end{tabular}
\end{table*}

\subsection{Comparative Evaluation of Systems}
\label{sec:comparative_evaluation}

In this section we compare the MKL based approach (where information from DCT-SA CPM and DCT-VQ CPM is fused)
against three recently proposed systems in
Wiliem~et~al.~\cite{Wiliem2013},
Cordelli~et~al.~\cite{Cordelli2011}
and Strandmark~et~al.~\cite{Strandmark2012}. 

The system in~\cite{Wiliem2013}, denoted {\it Wiliem},
is based on the DCT-SA descriptor with the DR spatial structure and Nearest Convex Hull classifier.
We denote the system in~\cite{Strandmark2012} by {\it Strandmark},
and used the code provided by the authors. 
The system employs various image statistics (eg.,~mean, standard deviation)
and morphological features (eg.,~number of objects, area). 
The random forest classifier is used.

We implemented the best reported descriptor in~\cite{Cordelli2011},
denoted by {\it Cordelli},
which is comprised of features such as image energy, mean and entropy, calculated from intensity and LBP channels.
The LBP channel is computed by computing the local pattern code for each pixel in the intensity channel.
We selected Logistic Boosting (LogitBoost) as the classifier instead of AdaBoost as the former obtained better performance.

The results are presented in Fig.~\ref{fig:system_comparisons}.
On the ICPRContest dataset, the Cordelli and Strandmark systems obtain comparable performance.
However, the performance of Cordelli is considerably lower than Strandmark on the SNPHEp-2 dataset,
indicating that the Cordelli system is not able to generalise to various recording conditions.
The Wiliem system obtains better performance than Cordelli and Strandmark on both datasets,
with a considerable advantage over Strandmark on SNPHEp-2.
However, the proposed MKL based system obtains the best performance on both datasets,
with a marked increase over Wiliem on the ICPRContest dataset.

\begin{figure*}[!tb]
  \centering
  \includegraphics[width=0.6\textwidth]{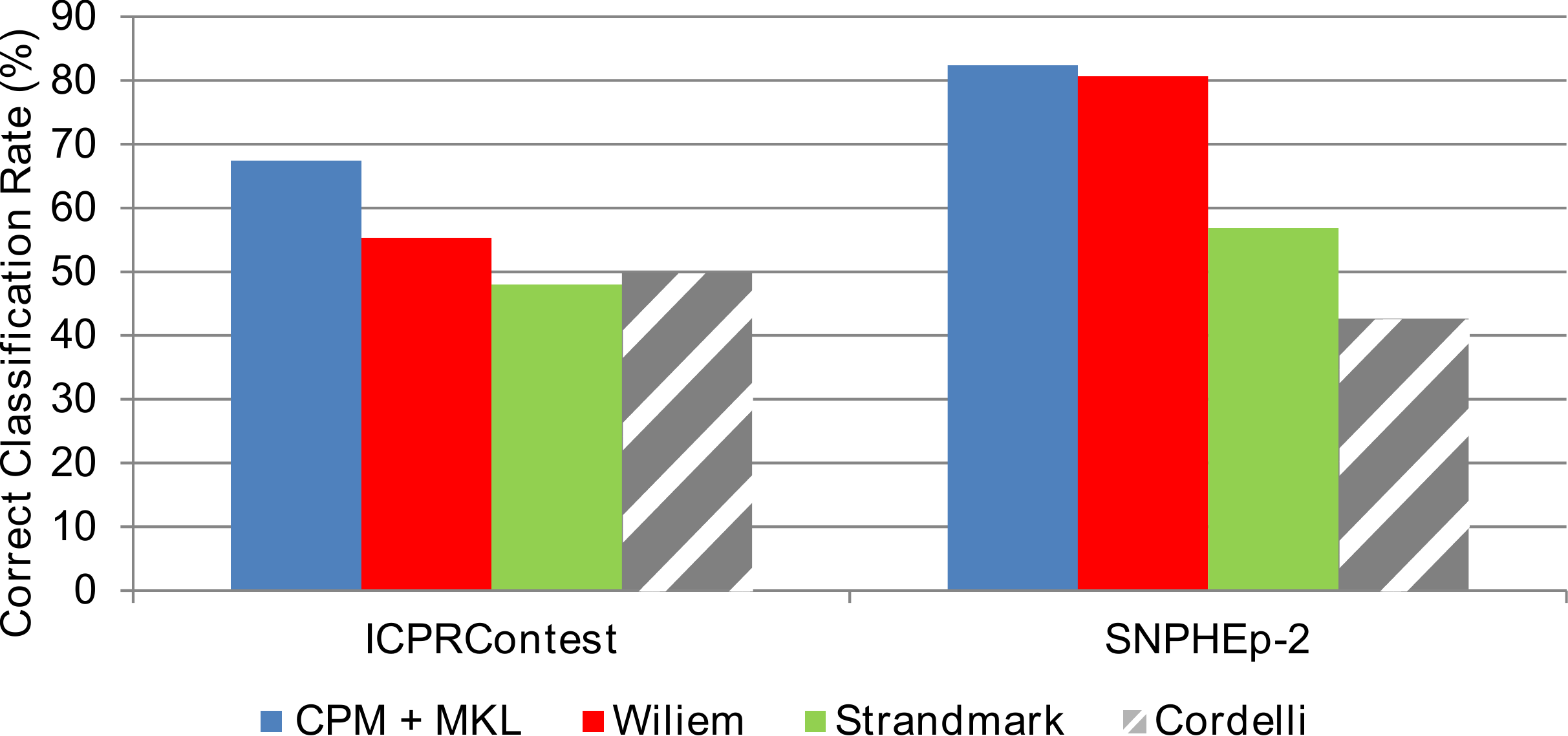} 
  \caption
    {
    Performance comparison of various systems on the ICPRContest and SNPHEp-2 datasets.    
    }
  \label{fig:system_comparisons}  
\end{figure*}

\newpage

\subsection{Cell Level and Image Level Performance on the ICPRContest Dataset}

Using the proposed MKL-based system from Section~\ref{sec:comparative_evaluation},
Fig.~\ref{fig:exp1_cell} shows the confusion matrix for the classification results on the ICPRContest dataset.
We also present the image level classification in Fig.~\ref{fig:exp1_image}. 
In image level classification we simply determine the label of an image based on the most frequent cell pattern.
In this setting, the MKL-based system achieves 71.4\%. 

We also report Leave-One-Out validation results for ICPRContest
in Table~\ref{tab:exp2_loo} as well as Figs.~\ref{fig:exp2_cell} and~\ref{fig:exp2_image}.
In this setting, the validation constructs 28 splits of train and test images,
where for each split cells belonging to a particular specimen image are used as the test images, and the rest as training images.

\begin{figure*}[!b]
  \centering
  \includegraphics[width=0.6\textwidth]{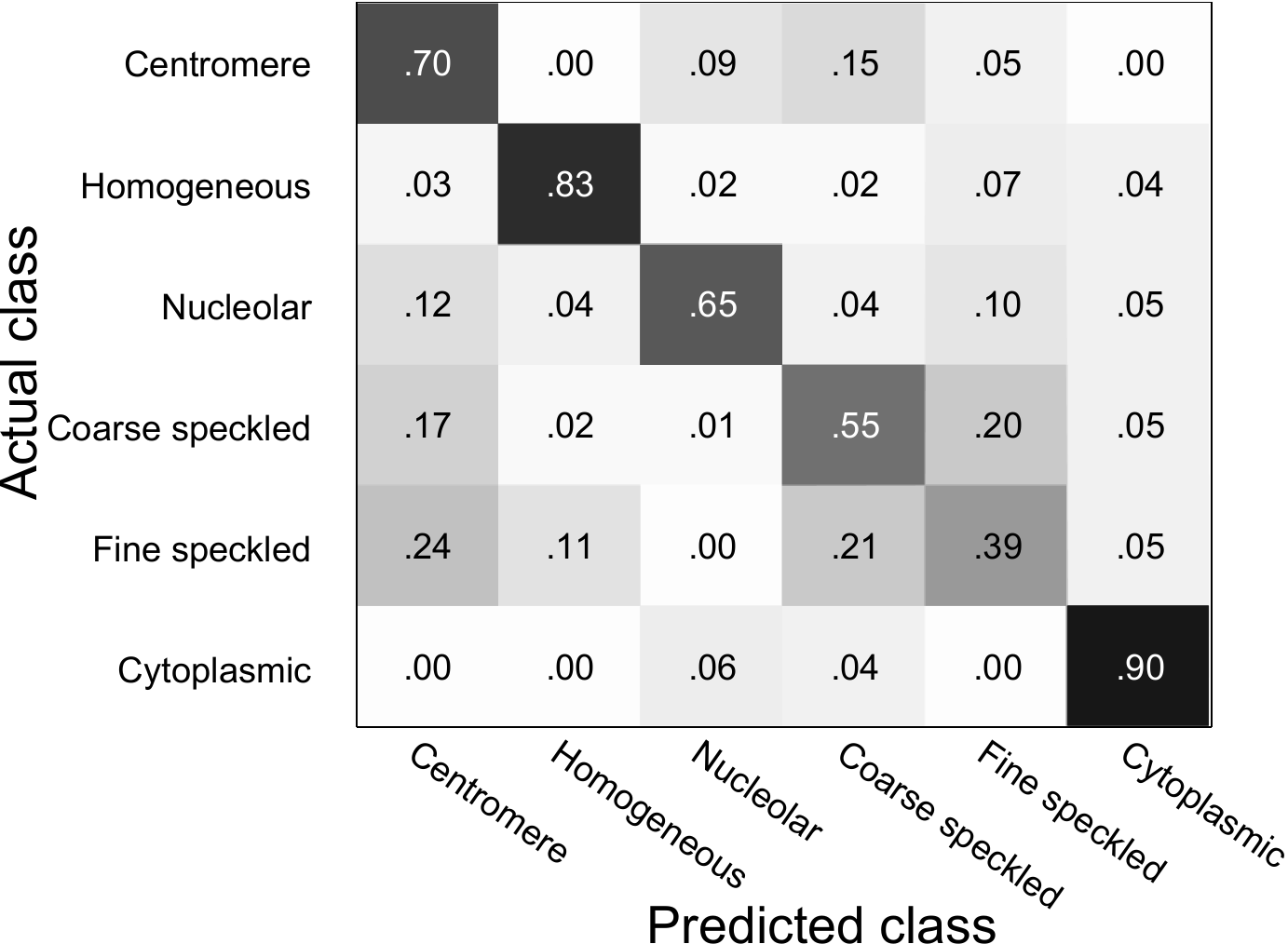} 
  \caption
    {
    Cell level confusion matrix of the proposed MKL-based system on the ICPRContest dataset.
    Each row and column represents instances of an actual class and predicted class, respectively.
    The elements in every row are normalised to one.
    The average accuracy is 67.4\%.
    Note that as the number of instances in each actual class is different,
    the average accuracy cannot be obtained by averaging the diagonal elements of the matrix.
    }
  \label{fig:exp1_cell}  
\end{figure*}

\begin{figure*}[!t]
  \centering
  \includegraphics[width=0.6\textwidth]{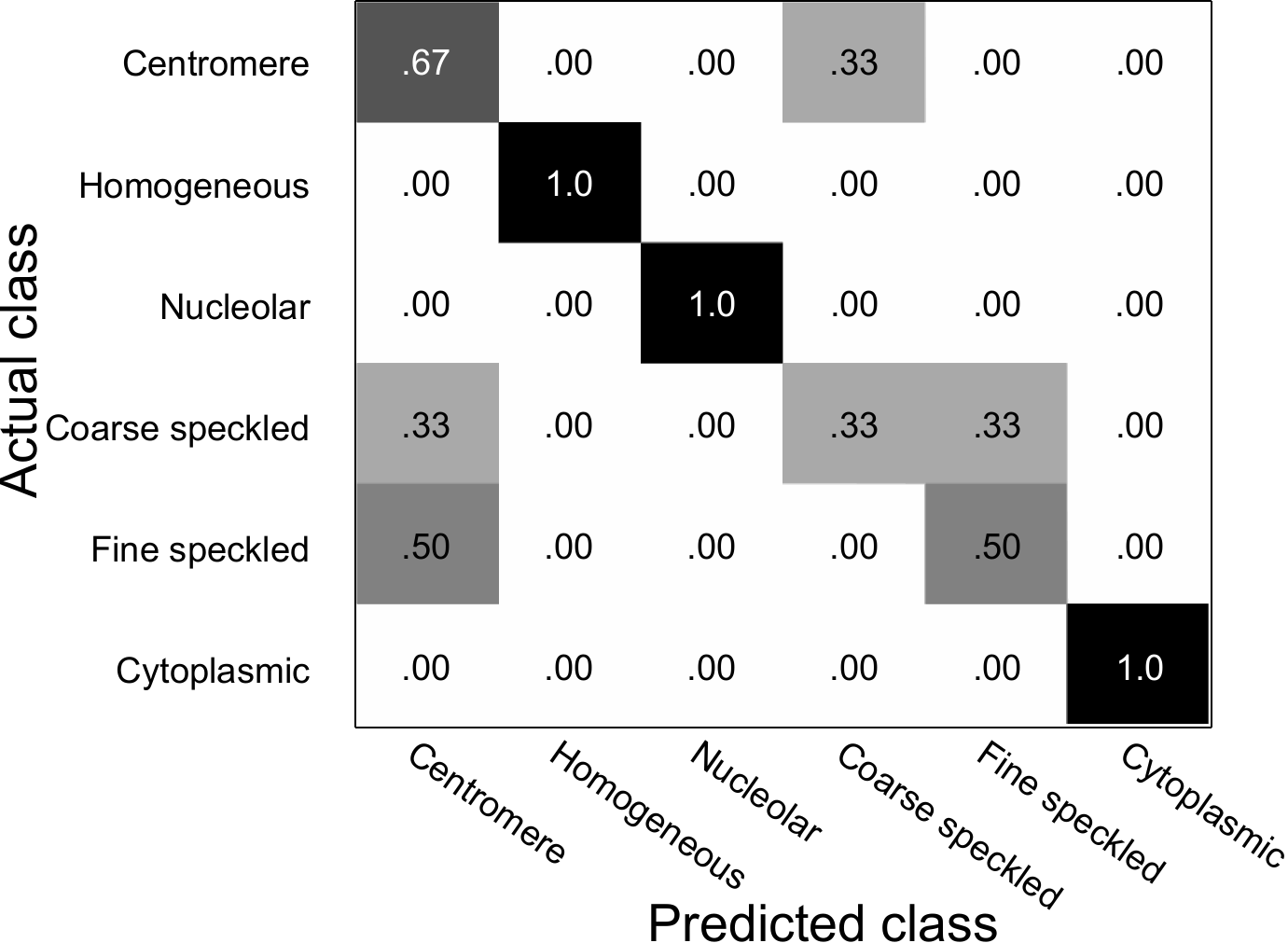} 
  \caption
    {
    Image level confusion matrix of the proposed MKL-based system on the ICPRContest dataset. The average accuracy is 71.4\%.
    }
  \label{fig:exp1_image}  
\end{figure*}

\begin{figure*}[!t]
  \centering
  \includegraphics[width=0.6\textwidth]{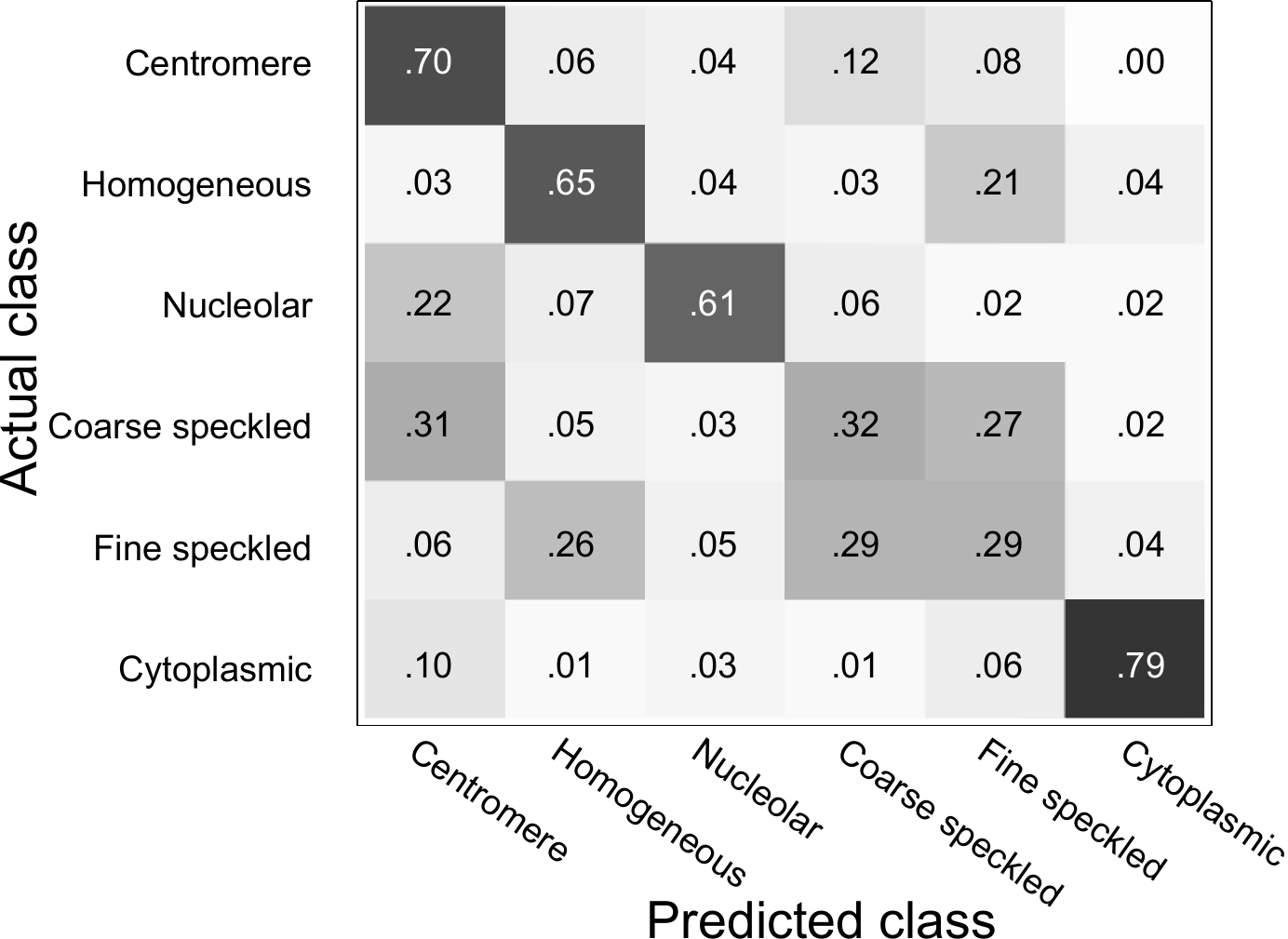} 
  \caption
    {
    Cell level confusion matrix of the proposed MKL-based system on the ICPRContest dataset using Leave-One-Out validation protocol. The average accuracy is 56.8\%.
    }
  \label{fig:exp2_cell}  
\end{figure*}

\begin{figure*}[!t]
  \centering
  \includegraphics[width=0.6\textwidth]{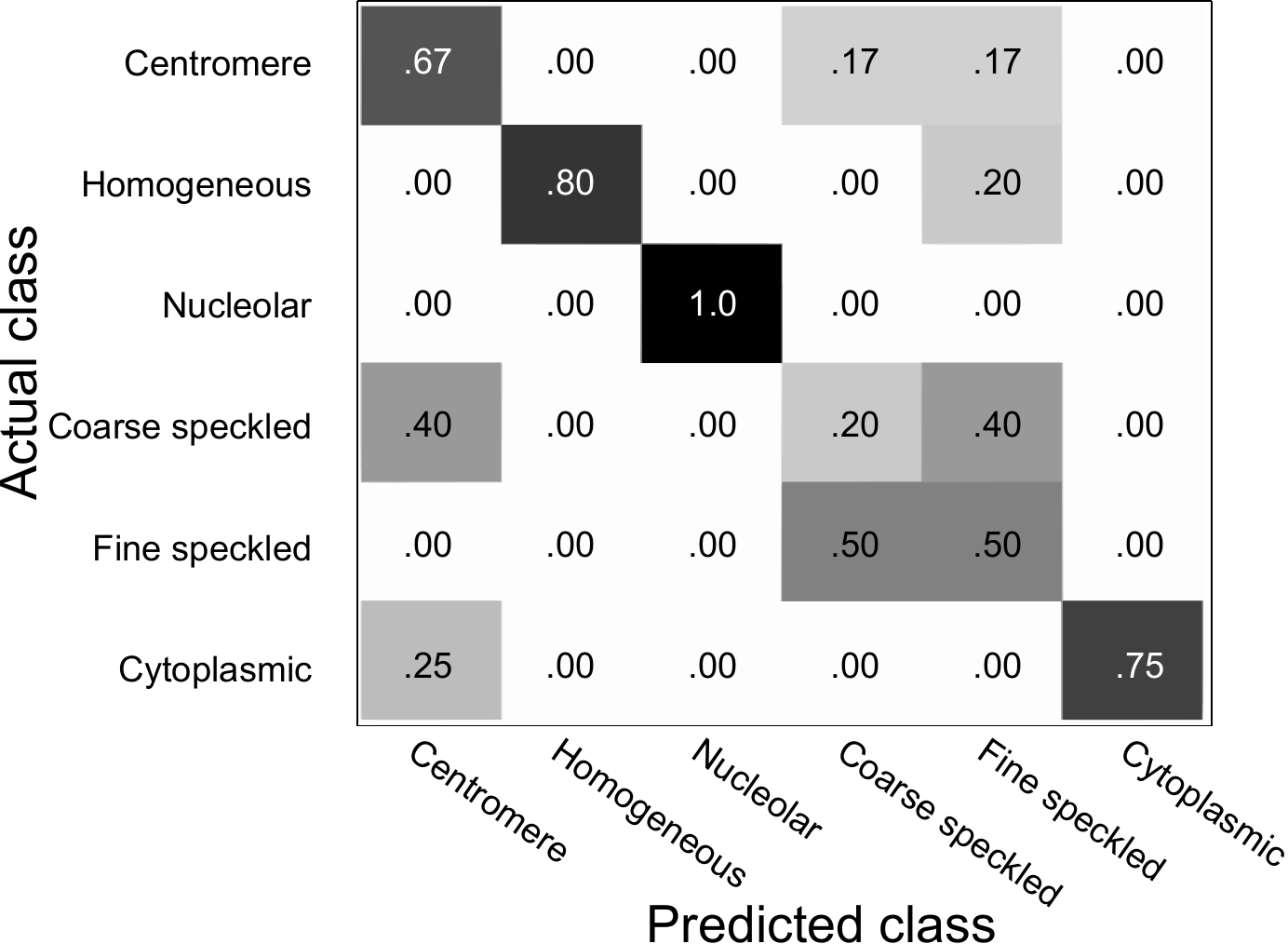} 
  \caption
    {
    Image level confusion matrix of the proposed MKL-based system on the ICPRContest dataset using Leave-One-Out validation protocol. The average accuracy is 64.3\%.
    }
  \label{fig:exp2_image}  
\end{figure*}

\begin{table*}[!t]
  \centering  
  \caption
    {
    Cell level classification performance for each cell image.
    Ce~=~Centromere;
    Ho~=~Homogeneous;
    Nu~=~Nucleolar;
    Co~=~Coarse speckled;
    Fi~=~Fine speckled;
    Cy~=~Cytoplasmic.
    }
  \label{tab:exp2_loo}
  \vspace{0.5ex}
  \begin{tabular}{c|l|cccccc|cccccc}
    \toprule
    \textbf{Specimen} & \textbf{True class} & \multicolumn{12}{c}{\textbf{Cells assigned to each class}}  \\                   
    \textbf{image}         &                     & \multicolumn{6}{c}{\textbf{(in absolute number)} }  & \multicolumn{6}{c}{\textbf{(in \%)}}    \\
    \textbf{number}        &                     & \textbf{Ce} & \textbf{Ho} & \textbf{Nu} & \textbf{Co} & \textbf{Fi} & \textbf{Cy}  & \textbf{Ce} & \textbf{Ho} & \textbf{Nu} & \textbf{Co} & \textbf{Fi} & \textbf{Cy}  \\
    \midrule
 1 & Homogeneous      & ~0  &  \textbf{43}  &  ~2  &  ~0  &  16  &  ~0 & ~0.0  &  \textbf{70.5}  &  ~3.3  &  ~0.0  &  26.2  &  ~0.0 \\
 2 & Fine speckled    & ~4  &  ~8  &  10  &  \textbf{19}  &  ~6  &  ~1 & ~8.3  &  16.7  &  20.8  &  \textbf{39.6}  &  12.5  &  ~2.1 \\
 3 & Centromere       & \textbf{87}  &  ~0  &  ~1  &  ~1  &  ~0  &  ~0 & \textbf{97.8}  &  ~0.0  &  ~1.1  &  ~1.1  &  ~0.0  &  ~0.0 \\
 4 & Nucleolar        & 23  &  ~3  &  \textbf{32}  &  ~0  &  ~5  &  ~3 & 34.8  &  ~4.5  &  \textbf{48.5}  &  ~0.0  &  ~7.6  &  ~4.5 \\
 5 & Homogeneous      & ~8  &  \textbf{26}  &  ~7  &  ~0  &  ~5  &  ~1 & 17.0  &  \textbf{55.3}  &  14.9  &  ~0.0  &  10.6  &  ~2.1 \\
 6 & Coarse speckled  & \textbf{44}  &  ~9  &  ~0  &  15  &  ~0  &  ~0 & \textbf{64.7}  &  13.2  &  ~0.0  &  22.1  &  ~0.0  &  ~0.0 \\
 7 & Centromere       & \textbf{54}  &  ~1  &  ~1 &  ~0   &  ~0  &  ~0 & \textbf{96.4}  &  ~1.8  &  ~1.8  &  ~0.0  &  ~0.0  &  ~0.0 \\
 8 & Nucleolar        & 25  &  ~0  &  \textbf{30}  &  ~0  &  ~0  &  ~1 & 44.6  &  ~0.0  &  \textbf{53.6}  &  ~0.0  &  ~0.0  &  ~1.8 \\
 9 & Fine speckled    & ~0  &  20  &  ~1  &  ~2  &  \textbf{23}  &  ~0 & ~0.0  &  43.5  &  ~2.2  &  ~4.3  &  \textbf{50.0}  &  ~0.0 \\  
 10 & Coarse speckled & \textbf{16}  &  ~0  &  ~0  &  ~7  &  10  &  ~0 & \textbf{48.5}  &  ~0.0  &  ~0.0  &  21.2  &  30.3  &  ~0.0 \\
 11 & Coarse speckled & ~1  &  ~0  &  ~1  &  ~7  &  \textbf{32}  &  ~0 & ~2.4  &  ~0.0  &  ~2.4  &  17.1  &  \textbf{78.0}  &  ~0.0 \\
 12 & Coarse speckled & ~4  &  ~0  &  ~5  &  \textbf{35}  &  ~0  &  ~5 & ~8.2  &  ~0.0  &  10.2  &  \textbf{71.4}  &  ~0.0  &  10.2 \\
 13 & Centromere      & 13  &  ~0  &  ~3  &  \textbf{30}  &  ~0  &  ~0 & 28.3  &  ~0.0  &  ~6.5  &  \textbf{65.2}  &  ~0.0  &  ~0.0 \\
 14 & Centromere      & ~4  &  19  &  ~1  &  9  &  \textbf{30}  &   ~0 & ~6.3  &  30.2  &  ~1.6  &  14.3  &  \textbf{47.6}  &  ~0.0 \\
 15 & Fine speckled   & ~9  &  ~7  &  ~0  &  \textbf{36}  &  ~3  &  ~8 & 14.3  &  11.1  &  ~0.0  &  \textbf{57.1}  &  ~4.8  &  12.7 \\
 16 & Centromere      & \textbf{36}  &  ~0  &  ~2  &  ~0  &  ~0  &  ~0 & \textbf{94.7}  &  ~0.0  &  ~5.3  &  ~0.0  &  ~0.0  &  ~0.0 \\
 17 & Coarse speckled & ~0  &  ~2  &  ~0  &  ~3  &  \textbf{14}  &  ~0 & ~0.0  &  10.5  &  ~0.0  &  15.8  &  \textbf{73.7}  &  ~0.0 \\
 18 & Homogeneous     & ~0  &  19  &  ~0  &  ~3  &  \textbf{14}   &  ~0 & ~0.0  &  45.2  &  ~0.0  &  7.1  &  \textbf{47.6}   &  ~0.0 \\
 19 & Centromere      & \textbf{56}  &  ~0  &  ~7  &  ~2  &  ~0  &  ~0 & \textbf{86.2}  &  ~0.0  &  10.8  &  ~3.1  &  ~0.0  &  ~0.0 \\
 20 & Nucleolar       & ~3  &  ~2  &  \textbf{41}  &  ~0  &  ~0  &  ~0 & ~6.5  &  ~4.3  &  \textbf{89.1}  &  ~0.0  &  ~0.0  &  ~0.0 \\
 21 & Homogeneous     & ~1  &  \textbf{50}  &  ~3  &  ~0  &  ~7  &  ~0 & ~1.6  &  \textbf{82.0}  &  ~4.9  &  ~0.0  &  11.5  &  ~0.0 \\
 22 & Homogeneous     & ~1  &  \textbf{78}  &  ~2  &  ~7  &  20  &  11 & ~0.8  &  \textbf{65.5}  &  ~1.7  &  ~5.9  &  16.8  &  ~9.2 \\
 23 & Fine speckled   & ~0  &  20  &  ~0  &  ~3  &  \textbf{28}  &  ~0 & ~0.0  &  39.2  &  ~0.0  &  ~5.9  &  \textbf{54.9}  &  ~0.0 \\
 24 & Nucleolar       & ~1  &  11  &  \textbf{45}  &  15  &  ~1  &  ~0 & ~1.4  &  15.1  &  \textbf{61.6}  &  20.5  &  ~1.4  &  ~0.0 \\
 25 & Cytoplasmic     & \textbf{10}  &  ~0  &  ~0  &  ~0  &  ~6  &  ~8 & \textbf{41.7}  &  ~0.0  &  ~0.0  &  ~0.0  &  25.0  &  33.3 \\
 26 & Cytoplasmic     & ~0  &  ~0  &  ~1  &  ~0  &  ~0  &  \textbf{33} & ~0.0  &  ~0.0  &  ~2.9  &  ~0.0  &  ~0.0  &  \textbf{97.1} \\
 27 & Cytoplasmic     & ~1  &  ~1  &  ~0  &  ~1  &  ~0  &  \textbf{35} & ~2.6  &  ~2.6  &  ~0.0  &  ~2.6  &  ~0.0  &  \textbf{92.1} \\
 28 & Cytoplasmic     & ~0  &  ~0  &  ~2  &  ~0  &  ~1  &  \textbf{10} & ~0.0  &  ~0.0  &  15.4  &  ~0.0  &  ~7.7  &  \textbf{76.9} \\

    \bottomrule
  \end{tabular}
\end{table*}

\clearpage

\section{Conclusions}
\label{sec:sec_conclusions}

In this paper we have proposed a cell classification system comprised of a Cell Pyramid Matching (CPM) descriptor combined with Multiple Kernel Learning.
The inspiration for the proposed CPM approach is drawn from Spatial Pyramid Matching (SPM) and Dual Region (DR) descriptors.
The major contributions of this study are:
(1)~proposing a more effective adapted version of SPM for cell images;
(2)~an extensive study on Bag-of-Words descriptor variants and various spatial structures.

We evaluated numerous configurations on two publicly available datasets:
ICPR \mbox{HEp-2} cell classification contest dataset and the new \mbox{SNPHEp-2} dataset.
We found that DCT patch-level features in conjunction with soft-assignment/probabilistic encoding of histograms
lead to the highest discrimination performance. 
We also found that the proposed CPM spatial layout is more effective than SPM and DR structures.
The proposed CPM also has an advantage of not having heuristic parameters and leads to a much shorter descriptor length.
The experiments show that the proposed system consistently delivered high performance
and is more robust than three recent CAD systems presented in \cite{Cordelli2011,Strandmark2012,Wiliem2013}.

\section*{Acknowledgements}

This research was partly funded by Sullivan Nicolaides Pathology, Australia and the Australian Research Council (ARC) Linkage Projects Grant LP130100230. 
NICTA is funded by the Australian Government as represented by the {\it Department of Broadband, Communications and the Digital Economy},
as well as the Australian Research Council through the {\it ICT Centre of Excellence} program.
NUS-ZJU Sensor-Enhanced Social Media (SeSaMe) Centre is supported by the Singapore
National Research Foundation under its International Research Centre @ Singapore Funding Initiative and administered by
the Interactive Digital Media Programme Office.

\renewcommand{\baselinestretch}{1.11}\small\normalsize

\bibliographystyle{ieee}
\bibliography{references}

\end{document}